\documentclass[twocolumn,showpacs,preprintnumbers,amsmath,amssymb,aps]{revtex4}
\usepackage{graphicx}
\begin{document}
\title{Quantum ratchet effect for vortices}
\author{J.~B. Majer, J. Peguiron, M. Grifoni, M. Tusveld and J.~E. Mooij}
\date{\today}
\affiliation{  Department of Nanoscience, Delft University of
Technology,
 Lorentzweg 1, 2628 CJ Delft, The Netherlands \\ }
\begin{abstract}
We have measured a quantum ratchet effect for vortices moving in a quasi-one-dimensional Josephson junction array. In this solid-state device the shape of the vortex potential energy, and consequently the band structure, can be accurately designed. This band structure determines the presence or absence of the quantum ratchet effect, as observed in the presented experiments. In particular,  asymmetric structures possessing only one band below the barrier do not exhibit current rectification at low temperatures and bias currents. The quantum nature of transport is also revealed in a universal/non-universal  power-law dependence of the measured voltage-current characteristics for samples without/with rectification.
\end{abstract}
\maketitle
A ratchet, i.e. an asymmetric periodic structure, yields the possibility to extract net particle flow from unbiased driving \cite{Reimann01}. The last twenty years have seen a large activity aimed at the theoretical understanding of the role of classical fluctuations on the rectification mechanism. Many experimental demonstrations of the classical ratchet effect have been reported ranging from physical to biological systems \cite{Special02}. In particular, solid-state semiconducting \cite{Lorke98} and superconducting devices \cite{Falo02,Sterck02} allow for a controlled design. In contrast, the understanding of the role of quantum noise is still at its infancy. On the one hand, the task of describing  the interplay among quantum fluctuations, unbiased driving and spatial potential asymmetry is formidable. The qualitatively new character of quantum ratchets was first pointed out  in \cite{Reimann97}. Current rectification and reversals in ac-driven ratchet  potentials were investigated only  recently \cite{Scheidl02,Grifoni02}. On the other hand, the lack of experimental realizations of quantum ratchets lies in the difficulty of fabrication of micro- or nano-sized  structures with controlled asymmetry. Rectification of quantum fluctuations has so far only been reported in triangularly-shaped semiconductor heterostructures \cite{Linke99}. Here a current reversal with decrease of temperature was observed as predicted in \cite{Reimann97}.

\begin{figure}
\includegraphics[width=7cm]{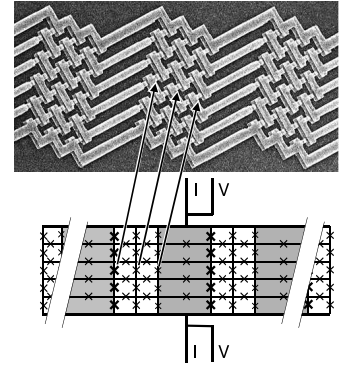}
\caption{
Strongly asymmetric array that exhibits ratchet effects. Top: scanning electron microscope picture. Bottom: schematic lay-out. Josephson junctions are represented by a cross, cells are areas enclosed by four junctions. All measured arrays have a length of 303 cells and a width of 5 cells between solid superconducting electrodes (busbars). Vortices are induced by an applied magnetic field perpendicular to the array. Cells  areas are 2.8 $\mu$m$^2$ (gray) and 0.7 $\mu$m$^2$ (white). Junctions indicated by arrows have areas of 240x100 nm$^2$, 200x100 nm$^2$ and 160x100 nm$^2$, respectively. Vortices have lower energy in cells with larger area and smaller junctions.
}
\label{RatchetSEM}
\end{figure}

In this letter, we report on the experimental observation of the quantum ratchet effect  for  vortices moving in quasi-one-dimensional Josephson junction arrays. Those arrays consist of a long, narrow network of Josephson junctions arranged in a rectangular lattice (Figs. \ref{RatchetSEM}, 2). A scanning electron microscope (SEM) picture of part of the most asymmetric mesoscopic device is shown in Fig. \ref{RatchetSEM}.
The potential shape felt by the vortices along the longitudinal direction can be accurately designed by properly choosing the junction sizes and/or the inter-junction distances. A quantum ratchet effect is observed in an asymmetric array providing a ratchet potential with three bands below the barrier. Strikingly, current rectification is {\em absent} in asymmetric arrays supporting only
 one band at low enough temperatures and bias currents.
 This is a consequence of time-reversal symmetry combined with translational properties of
 a one-band ratchet potential. 
In the following we introduce the basic properties of quasi-one-dimensional Josephson junctions array \cite{Bruder99}. Subsequently, we show how a ratchet potential for vortices can be designed. We finally report on the experimental results and their interpretation.

\begin{figure}
\includegraphics[width=9cm]{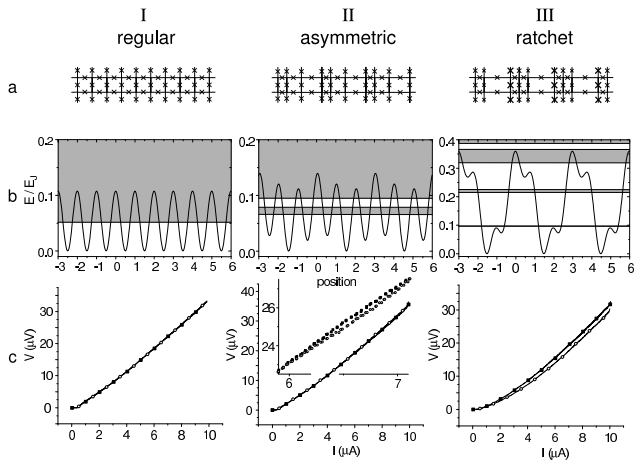}
\caption{
Samples and measurement results. The top row (a) indicates three supercells for each of the three arrays, row (b)
the resulting potential. The calculated vortex bands are also indicated. Sample I is a regular array. It has a cosine-shaped vortex potential and one energy band that connects with a continuum. Sample II shows a weak asymmetric
modulation on top of this cosine potential that leads to a gap in the spectrum with one band below the continuum. Sample III has a strongly asymmetric potential with three energy bands below the continuum. The bottom row (c) gives the measurement results for voltage (vortex current) versus bias current (vortex force), performed at
12~mK and a density $s$ of 0.61 vortices per supercell. Open circles label the positive branch and closed squares the negative branch. Sample III shows a clear asymmetry. Sample II shows a weak asymmetry only at high bias.
Inset: Blow-up of the V-I curve for sample II at large bias currents showing a weak ratchet effect.
}
\label{measurements}
\end{figure}

Our arrays consist of a network of rectangular superconducting islands, each weakly coupled to its four neighboring islands by Josephson junctions. Applying a magnetic field perpendicular to the array induces vortices in the system. The vortex density is proportional to the magnetic field strength \cite{Bruder99}. To confine the vortex motion in one dimension, superconducting strips (called busbars, cf. Fig. \ref{RatchetSEM}) are applied along the two long edges of the array. The superconducting current and voltage electrodes along the length of the array  repel the vortices, which consequently are forced to move along the centre row. When the vortices move, they create a voltage $V=sv\Phi_0/a$ across the array, which is measured at the end of the busbars. Here $s$ is the one-dimensional vortex density, $a$ is the average junction distance in the longitudinal direction, and $\Phi_0=h/2e$ is the superconducting flux quantum. Finally, $v$ is the average vortex velocity, and is influenced by all the microscopic details of the vortex dynamics. At low vortex densities, vortex-vortex interactions can be neglected \cite{note}, and the dynamics of a single vortex is homologous to that of a mass-carrying particle in a one-dimensional periodic potential \cite{Orlando91}. For a regular array the potential is approximately cosine-shaped \cite{Lobb83} (see Fig. \ref{measurements} b left).
\begin{figure}
\includegraphics[width=7cm]{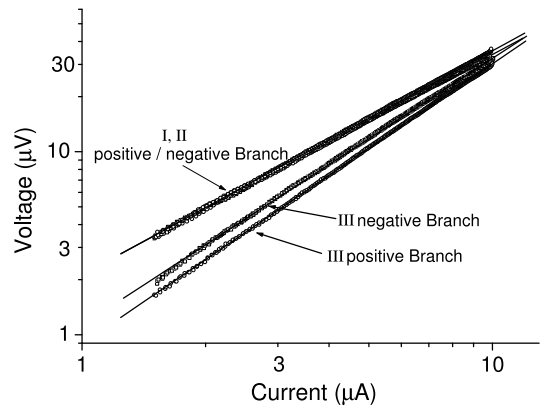}
\caption{
Power-law dependence of the $V-I$ characteristics at 12 mK. All the three samples exhibit above 1.5 $\mu$A a behavior $V\propto I^{\delta}$, $\delta >1$. Because three energy bands are involved in the dynamics,  sample III  shows a a larger power than samples I and II. The classical behavior would correspond to linear $V-I$ characteristics, i.e., $V\propto I$.} \label{logiv}
\end{figure}

The amplitude of the potential variation is proportional to the Josephson coupling energy $E_J=I_c \Phi_0/2\pi$, with $I_c$ the  critical current of a single junction. By varying the cell areas or the junction sizes the potential can be modified. Increasing the cell area increases the magnetic flux in the cell for a given magnetic field, which results in a lowering of the potential minimum. Modifying the junction size changes the Josephson energy and the height of the barrier for vortex motion between cells. These two methods allow for a tailored design of the potential, such as strength and symmetry. The mass of the vortex is proportional to the average capacitance $C$ of the junctions, $m_v\approx \Phi_0 2C/2a^2$  \cite{Lobb83,Orlando91}. The great advantage of a Josephson junction array is that both parameters, the potential $E_J$ and the mass $m_v$, can be controlled  by fabrication parameters. Damping is included phenomenologically upon introducing a friction term $-\eta v$ in the equations of motion. Within a resistively shunted junction (RSJ) model the viscosity of a regular array is estimated to be $\eta=\Phi_0^2/2a^2r_e$ \cite{Orlando91}. With  $r_e=1.5$k$\Omega$ being  the estimated  normal state resistance  of the junctions. Finally, vortices are put into motion upon injecting a current $I$ into the busbars, which exerts a Lorentz-like force on such particles. Such force  is  $F=I \Phi_0/N_c a$, where $N_c=304$ is the chosen number of junction columns in the array.

At low temperatures, tunneling processes through the barrier become relevant for a vortex of low mass and a weak pinning potential \cite{Oudenaarden1,Oudenaarden2}. In the junction arrays that we have designed the quantum regime is reached when the charging energy $E_c=e^2/2C$ is of the order of the amplitude of the potential, and for low enough temperatures. In this regime, the quantum vortex is able to tunnel, interfere or to localize \cite{Oudenaarden1,Oudenaarden2}. The degree of quantum coherence in the motion of a vortex is strongly dependent on the amount of dissipation. As a minimal model to describe the dissipative quantum dynamics of a single vortex, we consider the system-plus-bath Hamiltonian \cite{Weiss99}
$\hat H_{\rm tot} (t)=\hat H_{\rm R}+\hat x F(t)+\hat x\hat \xi(t)+ \hat H_{\rm B}$.
Here
\begin{equation}
\hat H_{\rm R}=\hat p^2/(2m_v)+V_{\rm R}(\hat x) \label{ratchet}
\end{equation}
is the isolated ratchet Hamiltonian for a vortex of mass $m_v$ moving in the periodic potential $V_{\rm R}(x+L)=V_{\rm R}(x)$. The action of the deterministic force is in the interaction term $\hat x F(t)$. Finally,  $\hat x\hat\xi+\hat H_{\rm B}$  is the standard Hamiltonian of an ensemble of harmonic oscillators
bi-linearly coupled to the vortex via the collective force
operator $\hat\xi$. The character of the bosonic bath is then
fully captured by the spectral function $J(\omega)=\eta\omega$,
being related to the
Fourier transform of the force-force correlator
$\langle\hat \xi(t)\hat\xi(s)\rangle$.
\begin{figure}
\includegraphics[width=7cm]{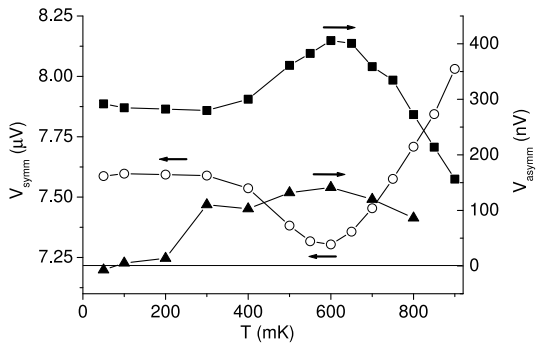}
\caption{Temperature dependence of the ratchet effect. Plotted with closed symbols  on the right scale (squares sample III, triangles sample II): $V_{\rm asymm}$ defined
as the difference between the voltages for negative and positive
currents. Plotted with open symbols (only sample III) on the left scale: $V_{\rm
symm}$ defined as the mean of these voltages. The bias
current is 6~$\mu$A for sample III, and 4~$\mu$A for sample II.} \label{tempdata}
\end{figure}

We have designed, fabricated and investigated three arrays with identical average properties (Fig. \ref{measurements}). Two of them (sample II and III) are superlattices, where a sequence of three cells is repeated along the length of the array. We refer to this set of three cells as the supercell with length $L$. The devices are fabricated from aluminum on a silicon substrate using shadow evaporation techniques. Sample I (regular) is an array with all cell areas equal to $a^2=1.4 \mu{\rm m}^2$, and all junction sizes equal to 100x200 ${\rm nm^2}$. These junctions have a capacitance $C$ of 2 fF. The critical current $I_c$ of the junctions is 210 nA, which is determined from the normal state resistance using the Ambegaokar-Baratoff relation \cite{Baratoff}. The  characteristic energy scales are $E_J \approx 10 E_c \approx 5$K$k_{\rm B}$. This regular sample serves as a reference for the other samples; the 'supercell' here consists of three identical basic cells. In sample II (weakly asymmetric) the cell size is varied periodically along the length of the array. The areas were chosen as 0.5-1-1.5 relative to the regular sample.
The five cells across the width of the array all have the same area. As expected, the resulting potential
(Fig. \ref{measurements} b centre) is asymmetric. The cell areas of the third sample (sample III, strongly asymmetric)
were chosen as 0.5-2-0.5 relative to the regular sample, and in addition the width of the vertical junctions
varied as 1.2-1-0.8 relative to the reference junction. The resulting potential (Fig. \ref{measurements} b right)
is strongly asymmetric. The three samples were fabricated on the same substrate under identical conditions. 
In the presented experiments a dc-current was swept between $-10\mu$A and $10 \mu$A.
We measured the {\em dc} voltage across the width of the array as a function of the applied bias current.
The ratchet effect manifests itself as a difference between the voltages for positive and negative bias currents.
To observe this difference we invert the negative branch of the current-voltage curve with respect to the origin
(Fig. \ref{measurements} c). 
Measurements  were carried out in a dilution refrigerator between 12~mK and 1~K.

The measurements at 12 mK show a clear ratchet effect for the strongly asymmetric sample.
 However,  no voltage asymmetry is observed between positive and negative current drives for the regular sample and for the weakly asymmetric sample at low bias currents. The symmetry for the regular array serves as a check for the validity of the experimental methods. The lack of  voltage asymmetry for sample II at low currents is remarkable, however. This can be understood by observing that the ratchet Hamiltonian for sample II has only one energy band that is well separated by a gap from the continuum at higher energies (Fig. \ref{measurements} b centre). Due to the low temperature (12~mK = 0.0024~$E_J$), only the low lying energy band ${\cal E}_1(k)$ is occupied. As we now show, despite the asymmetry, a single band ratchet Hamiltonian supports no current rectification.

The energy bands of a generic isolated ratchet Hamiltonian, Eq. (\ref{ratchet}), are obtained by solving the Schr\"odinger equation $\hat H_{\rm R}|n,k\rangle={\cal E}_n(k) |n,k\rangle$ ($n$ band index, $k$ wave vector). Time-reversal symmetry and the periodicity of the potential imply ${\cal E}_n(k)=E_n
+\sum_{m=1}^{\infty}(\Delta^{(m)}_n/2)\cos (mkL)$. To know wether
a single band asymmetric Hamiltonian supports a ratchet effect, we
restrict to the lowest band $n=1$ of the potential, and express
$\hat H_{\rm R}$ in the basis which diagonalizes the discrete
position operator $\hat x=\sum_{M=-\infty}^\infty
ML|M\rangle\langle M|$ (so termed discrete variable
representation, DVR, \cite{Thorwart00,note2}). $|M\rangle$ describes a state which is localized at cell $M$.
We observe that the driving and noise terms, $\hat x F$ and $\hat x\hat \xi$, are already diagonal in this basis.
Then the one-band Hamiltonian assumes the  expression
\begin{eqnarray}
\lefteqn{\hat H_{\rm
R}=\sum_{M=-\infty}^{\infty}(E_1|M\rangle\langle M| }
\nonumber\\&&+ \sum_{m=1}^{\infty}\frac{\Delta^{(m)}_1}{4}
(|M\rangle\langle M+m|+|M+m\rangle\langle M|))\;.
\label{ratchet-DVR}
\end{eqnarray}
Note that the same form holds  for a symmetric periodic potential. In fact a change  $\hat x\to -\hat x$ leaves $\hat H_{\rm R}$ invariant, and no ratchet effect occurs. The situation is different when more than one band
contributes to  transport. 
In sample II this occurs at large enough bias currents (inset Fig. 2c) or temperatures (Fig. 4). In Ref. \cite{Grifoni02} a detailed theory on the role of the higher bands has been developed  for few bands ac-driven quantum ratchets. This theory can also be adapted  to describe  the behavior of sample III in the region of moderate dc-bias currents ($FL\leq U_0$, with $U_0$ the potential barrier), such that Wannier-Stark states generated by the bands lying  above the barrier are not relevant.
A theory capable to describe the strong current regime is presently not available, and is
object of current research. 
  In general, when dissipative transitions between different energy bands occur, the total 
forward/backward rates  reflect the intra-well (vibrational motion) as well as the inter-well
(tunneling) dynamics. The breaking of detailed balance symmetry between backward and forward rates then implies a ratchet effect, i.e. $V(F)\neq -V(-F)$.

Another signature of quantum behavior is depicted in Fig. \ref{logiv}. For a classical dynamics and zero temperature  $V\propto I$ is expected above the critical current. However, all of the three samples exhibit a power-law behavior $V\propto I^\delta$ with exponent $\delta >1$ for moderate-to-large currents. Strikingly, for a large range of currents, sample I and II are lying on top of each other, despite the dissimilarity of the underlying potential.  
 For sample III we measure different powers for the two slopes which are higher than the powers of the previous samples.
We find it worth of notice, however, to observe that a power-law behavior $I\propto F^{2K-1}$, with $K$ the Kondo parameter, is expected to be seen at low temperatures ($k_{\rm B}T\ll FL$)  and moderate bias $F$ such that a tight-binding description is still applicable \cite{Weiss99} (it corresponds to  bias currents up to about  2$\mu$A).  We find it puzzling that at the largest bias currents of the experiments a similar power-law is also observed.

Figure \ref{tempdata} shows the temperature dependence of the
ratchet signal. On the right axis the difference $V_{\rm asymm}$
between the two branches is plotted for a fixed bias and density
 ($s=0.38$ sample II;   $s=0.28$ sample III). 
We also include the mean of the two branches $V_{\rm
symm}$ (plotted on the left axis) for sample III.  Below 350~mK down to the base temperature of 12~mK the
signals stay constant for sample III. 
This is a clear quantum signature.  In fact
a classical ratchet effect, resulting from thermal activation,
should disappear at low temperatures. Above 350~mK the ratchet
signal increases up to 650~mK and then decreases for higher
temperatures. The decrease  above 650~mK is due to
the reduction of the Josephson energy that sets in when the
critical temperature of the superconductor is approached, in
combination with the increase of the thermal energy. Due to the
weaker potential the asymmetry becomes less important and the
ratchet effect decreases. The fact that the mean transport 
increases is consistent
with that picture. The increase of the ratchet effect in the
intermediate regime (350~mK - 650~mK) is due to the generation of
quasi-particles. These quasi-particles are an additional source of
friction and cause the ratchet effect to increase
\cite{Reimann97}. Due to the additional damping the mean transport
is reduced.
Similar features are exhibited by sample II.

In summary, the band structure plays an important
role for a quantum ratchet. For an asymmetric
periodic potential with only one relevant energy band below the barrier
 the ratchet
effect is missing at low temperatures and bias. However
for a sample with three energy bands we measure a ratchet effect even 
at low temmperatures.
 Quantum signatures are also a saturation of the
signal at low temperatures, and a power-law behavior at moderate-to-high
currents. Finally,  additional friction increases
the ratchet effect.

We thank  M.~S. Ferreira, P.~Hadley, P.~H\"anggi, C.~J.~P.~M.
Harmans,  and M.~Thorwart for discussions; A. van den Enden and R.
Schouten for technical assistance. This work was supported by the
Dutch Foundation for Fundamental Research on Matter (FOM).


\begin{thebibliography}{99}
\bibitem{note}In our experiment all voltages depend linearly on the applied magnetic field for average flux up to 0.7 flux quanta per supercell. As the vortex density scales with the flux, this indicates a weak interaction among vortices.
\bibitem{note2} Note that the DVR basis
coincides with the Wannier basis only within a tight-binding
approximation. Information on the potential's asymmetry is in the
reflection properties of the DVR states.
\end{thebibliography}
\end{document}